\documentclass{article}
\usepackage{spconf,amsmath,graphicx}
\usepackage{amsfonts}
\usepackage{booktabs}
\usepackage{url}
\usepackage{cite}

\title{NEURAL NETWORK-BASED VIRTUAL MICROPHONE ESTIMATOR}
\name{Tsubasa Ochiai, Marc Delcroix, Tomohiro Nakatani, Rintaro Ikeshita, Keisuke Kinoshita, Shoko Araki}
\address{NTT corporation}
\begin{document}
\ninept
\maketitle
\begin{abstract}

Developing microphone array technologies for a small number of microphones is important due to the constraints of many devices.
One direction to address this situation consists of virtually augmenting the number of microphone signals, e.g., based on several physical model assumptions.
However, such assumptions are not necessarily met in realistic conditions.
In this paper, as an alternative approach, we propose a neural network-based virtual microphone estimator (NN-VME).
The NN-VME estimates virtual microphone signals directly in the time domain, by utilizing the precise estimation capability of the recent time-domain neural networks.
We adopt a fully supervised learning framework that uses actual observations at the locations of the virtual microphones at training time.
Consequently, the NN-VME can be trained using only multi-channel observations and thus directly on real recordings, avoiding the need for unrealistic physical model-based assumptions.
Experiments on the CHiME-4 corpus show that the proposed NN-VME achieves high virtual microphone estimation performance even for real recordings and that a beamformer augmented with the NN-VME improves both the speech enhancement and recognition performance.

\end{abstract}
%
\section{Introduction}
\label{sec:intro}

Microphone array signal processing \cite{benesty2008microphone}, which uses spatio-temporal information obtained with multiple microphones, has been an active research field for several decades and has been essential for the development of many applications such as noise reduction, source separation, and source localization.
The performance of array signal processing-based approaches depends heavily on the number of microphones available, and a sufficient number of microphones is required to achieve a high level of performance.
However, it is not always possible to equip commercial devices with many microphones due to structural constraints and cost restrictions.
For example, most smartphones have one or two microphones and only few high-end models may have up to four.
Therefore, developing array signal processing schemes that can operate with a small number of microphones is an important research topic. 

Recently, researchers \cite{katahira2016nonlinear} have proposed virtually increasing the number of microphones in an array by including virtual microphone signals generated by the interpolation between two real microphone observations.
They derived the phase component of a virtual microphone signal as a linear interpolation of the phases of real microphone signals, by introducing several assumptions such as a physical model of the plane wave propagation, W-disjoint orthogonality of the sources \cite{yilmaz2004blind}, and limiting array size small enough to avoid spatial aliasing.
The phase was combined with an estimate of the amplitude obtained, e.g., by $\beta$ divergence-based estimation \cite{katahira2016nonlinear}.
This approach could increase the number of microphones in an array and would improve the array processing performance in an under-determined condition.
However, this approach relies on the above strong assumptions that may not always hold in real conditions, and the method has only been tested on simulated array recordings.

In this paper, we propose an alternative approach to estimate virtual microphone signals based on a supervised learning framework using a time-domain neural network, which does not explicitly rely on the physical model assumptions.
Recently, neural networks operating directly in the time-domain, e.g., a time-domain audio separation network (TasNet) \cite{luo2018tasnet,luo2019conv}, have demonstrated a high level of performance for speech separation.
Moreover, it is confirmed in \cite{ochiai2020beam} that, when applying TasNet to microphone array recordings, the spatial information (i.e., phase) could be preserved, so that beamforming could be successfully constructed based on the output of TasNet.
These works have revealed the potential of neural networks to accurately estimate time-domain signals, i.e., predict both amplitude and phase information.
Motivated by these studies, in this paper, we investigated the potential of time-domain neural networks to estimate virtual microphone signals directly in the time-domain by utilizing the spatial information inferred from a few observed real microphone signals.
We call the method neural network-based virtual microphone estimator (NN-VME).

To enable the NN-VME to estimate the virtual microphone signals, we adopt a supervised learning framework, assuming that during training we have access to recordings at the location of the virtual microphone.
This assumption is reasonable in many circumstances, since we may have fewer constraints on the number of microphones during the system development (collection of training data) than during the actual deployment.
By adopting the supervised learning framework, we do not explicitly rely on the physical model assumptions, such as the plane wave propagation, unlike \cite{katahira2016nonlinear}.
Moreover, the neural network requires only microphone observations as training data.
Consequently, the network can be naturally trained and deployed on real recordings.

We tested the effectiveness of the proposed method both in terms of 1) virtual microphone estimation performance and 2) speech enhancement performance when combined with a beamformer.
To demonstrate the potential of the proposed method to deal with real recordings, we conducted these experiments with the well-known noisy speech benchmark, CHiME-4 corpus \cite{barker2015third}, which contains real recordings from noisy public environments.
Through the experiments, we confirm that the proposed NN-VME achieves high virtual microphone estimation performance and that a beamformer augmented with the proposed NN-VME improves the speech enhancement and automatic speech recognition (ASR) performance.

In this paper, as a typical use-case for the proposed method, we evaluate the combination of the NN-VME with beamforming for noise reduction.
Besides the combination of the NN-VME with array processing techniques, as the other potential application, the NN-VME could also be used for sound reproduction systems to generate virtual microphone signals at user-desired locations, where actual microphones could not be placed, e.g., due to physical constraints.


\section{PROPOSED METHOD: NEURAL NETWORK-BASED VIRTUAL MICROPHONE ESTIMATOR}

\subsection{Network architecture}

Figure~\ref{fig:nn-vme} shows the network architecture and the training procedure of the proposed NN-VME.
In the figure, for simplicity, the network receives two input channels corresponding to the observed real microphone, and it generates one output channel corresponding to the estimated virtual microphone.
In the following, we will explain the general case where we predict simultaneously more than one virtual microphone.

Let $\mathbf{r}_{c}$ be the $\mathcal{T}$-length time-domain waveform of the observed signal for the $c$-th real microphone, and $\hat{\mathbf{v}}_{c'}$ denotes the estimated signal for the $c'$-th virtual microphone.
Given the real microphone signals $\mathbf{r} = \{\mathbf{r}_{c=1}, \cdots, \mathbf{r}_{c=C_{\mathrm{r}}}\}$ as an input, the proposed NN-VME module estimates the virtual microphone signals $\hat{\mathbf{v}} = \{\hat{\mathbf{v}}_{c'=1}, \cdots, \hat{\mathbf{v}}_{c'=C_{\mathrm{v}}}\}$ as:
\begin{align}
\hat{\mathbf{v}} = \text{NN-VME}\bigl( \mathbf{r} \bigr),
\label{eq:nn-vme}
\end{align}
where, $C_{\mathrm{r}}$ denotes the number of observed channels, i.e., real microphones, and $C_{\mathrm{v}}$ denotes the number of virtually estimated channels, i.e., virtual microphones, and $\text{NN-VME}(\cdot)$ is a neural network.

We adopt an architecture for $\text{NN-VME}(\cdot)$ inspired by Conv-TasNet \cite{luo2019conv}, as it was shown to be able to estimate time-domain signals with high accuracy for speech separation.
Similar to the original Conv-TasNet, the network is mainly composed of a 1d-convolution encoder layer, internal convolution blocks, and a 1d-deconvolution decoder layer.
First, the encoder layer directly maps the time-domain signals to an intermediate representation, and then the intermediate representation is further processed by several convolution blocks.
Finally, the decoder layer converts the intermediate representation back to time-domain signals.

Note that the original TasNet estimates the {\it separated signals} at the location of the real microphone, while the proposed NN-VME estimates the {\it observed signals} at the location of the virtual microphone.

\subsection{Supervised training}

The proposed NN-VME adopts the supervised learning framework in order to enable the NN-VME module to estimate the virtual microphone signals.
Thus, in the training, we use actual microphone signals at the location of the virtual microphones as training targets.

\begin{figure}[t]
  \centering
  \includegraphics[width=0.68\linewidth]{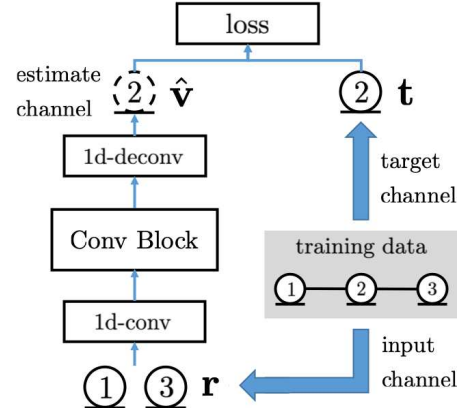}
\vspace{-2mm}
  \caption{Overview of network architecture and training procedure of neural network-based virtual microphone estimator}
\vspace{-3mm}
  \label{fig:nn-vme}
\end{figure}

For supervised training of the proposed NN-VME, we assume that a set of input and target signals $\{\mathbf{r}, \mathbf{t}\}$ is available.
Here, $\mathbf{t} = \{\mathbf{t}_{c'=1}, \cdots, \mathbf{t}_{c'=C_{\mathrm{v}}}\}$ and $\mathbf{t}_{c'}$ denotes the target signal for the $c'$-th virtual microphone.
Figure~\ref{fig:nn-vme} illustrates this situation; a subset of the microphones (e.g., channels $1$ and $3$) is assigned as network input $\mathbf{r}$, and another subset (e.g., channel $2$) is used as network target $\mathbf{t}$.

We train the network based on the time-domain loss between the estimated and actual signals at the location of the virtual microphones.
As the training loss, we adopt the scale-dependent signal-to-noise ratio (SNR)~\cite{roux2019sdr} as follows:
\begin{align}
\mathcal{L} = \sum_{c'=1}^{C_{\mathrm{v}}} 10 \log_{10} \biggl( \frac{\| \mathbf{t}_{c'} \|^{2}}{\| \mathbf{t}_{c'} - \hat{\mathbf{v}}_{c'} \|^{2}} \biggr),
\end{align}
where $\hat{\mathbf{v}} = \text{NN-VME}\bigl( \mathbf{r} \bigr)$, as described in Eq.~\eqref{eq:nn-vme}.

Note that, in contrast to the speech separation task, we simply need to estimate the virtual microphone signals and there is no permutation ambiguity, and thus permutation invariant loss is not necessary.
Unlike the training for the speech enhancement techniques, the training for the proposed method does not require parallel noisy and clean signals and it only requires the multi-channel noisy observation signals, which would be relatively easy and low-cost to prepare.
That is, the proposed model can be trained on real recordings, not simulated ones.
By exploiting a large amount of training data, we expect that the powerful neural network will be able to provide fine modeling of real recordings.

\section{APPLICATION EXAMPLE: BEAMFORMING WITH VIRTUAL MICROPHONE ESTIMATOR}

The proposed NN-VME enables the generation of virtual microphone signals, and thus could be used with various array processing techniques.
In this paper, to confirm the validity of the estimated virtual microphone signals for noise reduction applications, we investigated combining the NN-VME with a frequency domain beamformer.

\subsection{General procedure}


First, using the proposed NN-VME, we estimate the virtual microphone signals $\hat{\mathbf{v}} \in \mathbb{R}^{\mathcal{T} \times C_{\mathrm{v}}}$ given the real microphone signals $\mathbf{r} \in \mathbb{R}^{\mathcal{T} \times C_{\mathrm{r}}}$, as described in Eq.~\eqref{eq:nn-vme}, and obtain the {\it augmented microphone signals} $\mathbf{y} = [\mathbf{r}, \hat{\mathbf{v}}] \in \mathbb{R}^{\mathcal{T} \times C}$, where $C = C_{\mathrm{r}}+C_{\mathrm{v}}$.
Then, we apply a frequency-domain beamformer on top of the augmented microphone signals in frequency-domain representation, i.e., short-time Fourier transform (STFT), to obtain the enhanced speech signal.
Finally, we reconstruct the enhanced time-domain waveform using the inverse STFT.

The enhanced speech signal in the STFT domain $\hat{X}_{t,f} \in \mathbb{C}$, is obtained as $\hat{X}_{t,f} = \mathbf{w}_{f}^{\mathrm{H}} \mathbf{Y}_{t,f}$.
Here, $\mathbf{Y}_{t,f} \in \mathbb{C}^{C}$ is a vector comprising the $C$-channel STFT coefficients of the augmented microphone signals at time-frequency bin $(t,f)$, $\mathbf{w}_{f} \in \mathbb{C}^{C}$ is a vector comprising the beamforming filter coefficients, and $\empty^{\mathrm{H}}$ represents the conjugate transpose.

\subsection{Mask-based MVDR Formalization}

In this paper, we adopt the minimum variance distortionless response (MVDR) formalization of \cite{souden2009optimal}, and we compute the time-invariant filter coefficients $\mathbf{w}_{f}$ as follows:
\begin{align}
\mathbf{w}_{f} = \frac{(\mathbf{\Phi}^{\mathrm{N}}_{f})^{-1} \mathbf{\Phi}^{\mathrm{S}}_{f}}{\mathrm{Tr}((\mathbf{\Phi}^{\mathrm{N}}_{f})^{-1} \mathbf{\Phi}^{\mathrm{S}}_{f})} \mathbf{u},
\end{align}
where $\mathbf{\Phi}^{\mathrm{S}}_{f} \in \mathbb{C}^{C \times C}$ and $\mathbf{\Phi}^{\mathrm{N}}_{f} \in \mathbb{C}^{C \times C}$ are the spatial covariance (SC) matrices for speech and noise signals, respectively.
$\mathbf{u} \in \mathbb{R}^{C}$ is a one-hot vector representing the reference microphone.

Based on \cite{heymann2016neural}, we approximately estimate the SC matrices using time-frequency masks as follows:
\begin{align}
\mathbf{\Phi}^{\nu}_{f} &= \frac{1}{\sum_{t=1}^{T} m_{t,f}^{\nu}} \sum\limits_{t=1}^{T} m_{t,f}^{\nu} \mathbf{Y}_{t,f} \mathbf{Y}_{t,f}^{\mathrm{H}}, \label{eq:sc}
\end{align}
where $\nu \in \{\mathrm{S}, \mathrm{N}\}$.
$m_{t,f}^{S} \in [0, 1]$ and $m_{t,f}^{N} \in [0, 1]$ are the time-frequency masks for speech and noise, respectively.

\subsection{Virtual Microphone Loading}
\label{sec:loading}

Our preliminary experiments showed that while use of virtual microphones in beamforming was effective for increasing signal-to-distortion ratios (SDR) \cite{vincent2006performance}, it does not necessarily contribute to improving the ASR performance.
This is probably because processing artifacts are introduced by the virtual microphone estimation.
To reduce the effect of such artifacts, we introduced a virtual microphone loading term $\mathbf{Z} \in \mathbb{R}^{C}$ to the noise SC matrix $\mathbf{\Phi}^{\mathrm{N}}_{f}$:
%
\begin{equation}
\mathbf{\Phi}^{\mathrm{N}}_{f}   \leftarrow \mathbf{\Phi}^{\mathrm{N}}_{f} + \epsilon \mathbf{Z},
\label{eq:loading}
\end{equation}
where $\mathbf{Z} = \{z_{c,c'}\}_{c=1,c'=1}^{C,C}$ is a matrix of zeros excepts for the diagonal elements corresponding to the virtual microphone, i.e.,  $z_{c_{\mathrm{v}},c_{\mathrm{v}}} = 1$, $c_{\mathrm{v}}$ denotes the channel index corresponding to the virtual microphone, and $\epsilon$ is a loading hyperparameter that controls the contribution of the virtual microphone in the construction of the beamformer.
For example, when we set a larger value to $\epsilon$, it means that the virtual microphone is contaminated by larger noise that is not correlated with other microphones.
As a result, the estimated beamformer puts less weight on the virtual microphone channels.

\section{RELATION TO PRIOR WORK}

Our work is related to \cite{katahira2016nonlinear,yamaoka2017performance,jinzai2018microphone,yamaoka2019cnn}, which proposed to generate virtual microphone signals by interpolation/extrapolation of signals observed at real microphones.
In these methods, the phase components are estimated based on several physical model assumptions, which would make it difficult to use in practice.
For example, for the CHiME-4 corpus, the microphone spacing is too large to assume no spatial aliasing and thus it cannot be applied to our experimental settings.

In \cite{yamaoka2019cnn}, a supervised learning-based framework is also adopted to estimate the amplitude components of the virtual microphone.
This framework minimizes the loss function between the output of the beamformer and the clean signal, but it causes overfitting problem on the open test set.
In contrast, our work assumes the availability of recordings at the location of the virtual microphone during training, and it minimizes the loss function between the output of the NN-VME and actual microphone signals at the locations.
We experimentally confirmed the generalization capability for the open test set and even for real recordings as shown in the experimental results of Section~\ref{sec:exp}.


\section{EXPERIMENT}
\label{sec:exp}
To evaluate the proposed NN-VME, we conducted two types of evaluations: 1) evaluation of the virtual microphone estimation performance by the proposed NN-VME and 2) evaluation of the enhancement performance by the beamformer with the estimated virtual microphone.
In the experiment, we report the results that estimate one virtual microphone, but our method can be naturally extended to estimate multiple virtual microphones.

\subsection{Experimental conditions}

\begin{figure}[t]
  \centering
  \includegraphics[width=0.5\linewidth]{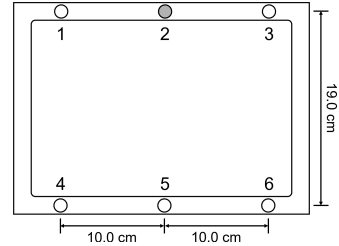}
  \caption{Microphone array geometry for CHiME-4 corpus. All microphones face forward except for microphone 2.}
\vspace{-3mm}
  \label{fig:tablet}
\end{figure}

We evaluated the proposed NN-VME on the CHiME-4 corpus \cite{barker2015third}.
The CHiME-4 corpus consists of speech recorded using a tablet device equipped with a rectangular microphone array with 6 channels, as illustrated in Figure~\ref{fig:tablet}.
The corpus contains not only simulated data but also real recordings from noisy public environments.

The training set consisted of 3 hours of real speech data uttered by 4 speakers and 15 hours of simulation speech data uttered by 83 speakers.
The evaluation set consisted of 1320 utterances of real and simulated noisy speech data uttered by 4 speakers, respectively.
Out of these utterances, the CHiME-4 organizers reported that 12\% of all the real data were affected by microphone failures, mainly for channels 4 and 5 \cite{chime1}.
Dealing with such microphone failures is out of the scope of this study.
Therefore, to remove the utterances with microphone failures, we excluded microphone signals whose minimum cross-correlation score among channels 4, 5, and 6 was less than $0.9$, using the cross-correlation coefficients provided by the organizers \cite{chime2}.
The resultant evaluation set consisted of 1149 utterances (i.e., approximately 13\% of the data were rejected).

As the evaluation metrics, we used the SDR of BSSEval \cite{vincent2006performance} and the word error rate (WER).
To evaluate the virtual microphone estimation performance, we computed the SDR between the estimated virtual microphone signals and the observed real microphone signals at the channel corresponding to the virtual microphone.
To evaluate the enhancement performance of the beamformer, we used the clean reverberant signals at the fourth channel as reference.
Since we required access to clean signals, this evaluation could only be performed on the simulated data.
We used Kaldi's CHiME-4 recipe \cite{povey2011kaldi,kaldi} to evaluate the ASR performance, which consisted of a deep neural network-hidden Markov model hybrid acoustic model \cite{bourlard1994connectionist,hinton2012deep} trained with the lattice-free maximum mutual information criterion \cite{povey2016purely}.
We used a trigram language model for decoding.

\subsection{Experimental configurations}

We adopted the Conv-TasNet-based network architecture for the network configuration of our proposed NN-VME.
By following the notations of \cite{luo2019conv}, we set the hyperparameters as follows: $N = 256$, $L = 20$, $B = 256$, $H=512$, $P=3$, $X=8$, and $R=4$.

We trained the proposed NN-VME using both simulated and real data of the training set, by adopting the Adam algorithm~\cite{kingma2015adam} with an initial learning rate of 0.0001 and the gradient clipping~\cite{pascanu2013difficulty}.
We stopped the training procedure after 200 epochs.

For the MVDR beamformer, we used the trained mask estimation model \cite{heymann2016neural} provided in the GitHub repository \cite{nngev}, which was used in the Kaldi's CHiME-4 recipe.
For the STFT computation, we used a Blackman window with a length and shift set at 64 ms and 16 ms, respectively.
In the ASR experiment, we set the loading hyperparameter $\epsilon$ of Eq.~\eqref{eq:loading} to $0.05$.

\subsection{Experimental results}

\subsubsection{Evaluation of virtual microphone estimation performance}

\begin{table}[t]
  \renewcommand{\arraystretch}{1.0}
  \caption{SDR [dB] for virtual microphone estimator, in which noisy observed signal is used as reference signal}
\vspace{3mm}
  \label{tab:result1}
  \centering
  \scalebox{0.95}{
  \begin{tabular}{ c l c c c }
    \toprule
    mic type & eval ch & ref ch & simu & real \\
    \midrule
    RM & 4 & 5 & 12.1 & 8.8 \\
    VM & 5 (4,6) & 5 & 16.6 & 13.8 \\
    VM & 5 (3,4,6) & 5 & 16.5 & 13.8 \\
    \midrule
    RM & 5 & 6 & 8.3 & 7.8 \\
    VM & 6 (4,5) & 6 & 12.3 & 11.8 \\
    VM & 6 (3,4,5) & 6 & 12.5 & 11.9 \\
    \bottomrule
  \end{tabular}
  }
\vspace{-2mm}
\end{table}

Table~\ref{tab:result1} shows the SDR scores of the evaluated NN-VME on the simulated and real recordings, where {\it RM} refers to real microphone, while {\it VM} refers to virtual microphone estimated by the proposed NN-VME.
Here, note that the reference signal for the SDR computation is the noisy observation signal at the channel corresponding to the virtual microphone rather than the clean signal, and thus the virtual microphone estimation performance can be evaluated even for the real recordings.

In the table, the first column  (``eval ch'') shows the channel index of the virtual or real microphone signal used as the estimated signal in the SDR calculation. The second column (``ref ch'') shows the channel index of the real microphone signal used as the reference signal.
Here, the notation ``5 (4,6)'' indicates that the virtual microphone signal at channel 5 was estimated using the real microphone signals at channels 4 and 6.
As a baseline, we compare the scores with the SDR obtained with the closest real microphones (i.e., the closest in terms of SDR).
These results are shown in the first (eval ch 4, ref ch 5) and fourth rows (eval ch 5, ref ch 6) of Table~\ref{tab:result1}.

Table~\ref{tab:result1} shows that the estimated signals by the proposed NN-VME modules (e.g., ``$5\ (4,6)$'') achieved much higher SDR scores than the observed signal recorded at the close microphone (e.g., ``$4$'').
These results demonstrate that, even for real recordings, the proposed NN-VME, i.e., the supervised learning-based virtual microphone estimation framework, has a potential to estimate the virtual microphone signals, which are not really observed by a microphone, by utilizing the spatial information inferred from a few observed real microphone signals. 

The table shows results for interpolation, i.e., the virtual microphone is located between the real microphones (e.g., ``$5\ (4,6)$''), and extrapolation in the horizontal direction (e.g., ``$6\ (4,5)$'').
In both cases, the NN-VME could predict the virtual microphone with an SDR of more than about 12 dB.
In addition, we observed  that increasing the number of the input channels (e.g., ``$5\ (3,4,6)$'') did not contribute to a significant improvement of the virtual microphone estimation performance.
Investigating the behavior of the proposed method for various microphone array configurations is an interesting direction for future work.


\subsubsection{Evaluation of beamformer enhancement performance}

Table~\ref{tab:result2} shows the SDR scores of the evaluated beamformers on the simulated data.
Here, {\it VM BF} refers to the beamformer with the estimated virtual microphone, while {\it RM BF} refers to the beamformer only with the real microphone.
In the table, the columns ``real'' and ``virtual'' in ``used ch'' denote the channel indices corresponding to the real and virtual microphones, which are used for constructing the beamformer, respectively.
For example, the ``VM BF'' of row (4) is constructed using the two real microphone signals (i.e., channels $4$ and $6$) and one virtual microphone signal (i.e., channel $5$).


Table~\ref{tab:result2} shows that the proposed VM BF (e.g., row~(4)) successfully achieved higher SDR score compared to the RM BF constructed with the same real microphone signals (e.g., row~(2)).
Here, another RM BF (e.g., row~(3)) would correspond to the upper-bound performance of the VM BF.

In addition to the above SDR-based evaluation, we conducted an ASR evaluation to evaluate the beamformer's performance on the real recordings.
Table~\ref{tab:result2} also shows the WER of the evaluated RM and VM BFs on the real data.
From the table, we confirmed that, even for the real recordings, the proposed VM BF (e.g., row (4)) reduced the WER compared to its corresponding RM BF (e.g., row (2)) by up to 0.9 \%.
A similar trend is observed when using more microphones (i.e., rows (5)-(7)).

These results demonstrate that the estimated virtual microphone signals can contribute to improved enhancement performance when combined with a beamformer.

In the table, the results of the VM BF with the virtual microphone loading, as described in Section~\ref{sec:loading}, are reported.
The WER scores of the VM BF without the loading are 15.0 \% for row (4) and 13.4 \% for row (7), respectively.
This confirms the effectiveness of the virtual microphone loading technique to improve the ASR performance for the VM BFs.


\begin{table}[t]
  \renewcommand{\arraystretch}{1.0}
  \caption{SDR [dB] (higher is better) and WER [\%] (lower is better) for beamformer, in which clean signal is used as reference signal}
\vspace{3mm}
  \label{tab:result2}
  \centering
  \scalebox{0.95}{
  \begin{tabular}{ l l c c c }
    \toprule
    Method & \multicolumn{2}{c}{used ch} & SDR & WER\\
    & real & virtual & (simu) & (real)\\
    \midrule
    (1)\ \ no process &\ \ - & - & 8.6 & 15.8\\
    \midrule
    (2)\ \ RM BF & 4,6 & - & 10.8 & 12.0\\
    (3)\ \ RM BF & 4,5,6 & - & 14.2 & 9.4\\
    (4)\ \ VM BF & 4,6 & 5 & 13.4 & 11.1\\
    \midrule
    (5)\ \ RM BF & 3,4,6 & - & 12.7 & 10.0\\
    (6)\ \ RM BF & 3,4,5,6 & - & 15.2 & 8.5\\
    (7)\ \ VM BF & 3,4,6 & 5 & 14.2 & 9.5 \\
    \bottomrule
  \end{tabular}
  }
\vspace{-2mm}
\end{table}


\section{CONCLUSION}

This paper proposed a novel virtual microphone estimation scheme, i.e., NN-VME, that employs a time-domain neural network architecture to directly predict the waveform of a virtual microphone signal.
The proposed method relies on supervised learning framework and the high modeling capability of the network to build a model that can accurately predict virtual microphone signals based on a few real microphone observations.

We showed experimentally that the proposed NN-VME could precisely predict virtual microphone signals even for real recordings, and that these predicted signals could help boost beamforming performance.

Future work will include, evaluation of the effectiveness of the proposed method for 1) various microphone array configurations, 2) various acoustic conditions, and 3) other array processing techniques, e.g., source separation \cite{hyvarinen2000independent} and source localization \cite{schmidt1986multiple}.

\vfill
\pagebreak

\bibliographystyle{IEEEbib}
\bibliography{refs}

\end{document}